\documentclass[reprint,groupedaddress,prb]{revtex4-2}
\usepackage{amsmath}

\usepackage{graphicx}% Include figure files
\usepackage{dcolumn}% Align table columns on decimal point
\usepackage{bm}%
\usepackage{braket}

\usepackage{hyperref}
\hypersetup{
    colorlinks=true,
    linkcolor=blue,
    filecolor=blue,       
    urlcolor=blue,
    citecolor=blue
}

\begin{document}

\title{Partial Landau-Zener transitions and applications to qubit shuttling}

\author{Jonas R. F. Lima}
%\email[]{jonas.de-lima@uni-konstanz.de}
\author{Guido Burkard}
%\email[]{guido.burkard@uni-konstanz.de}

%\homepage[]{Your web page}
%\thanks{}
\affiliation{Department of Physics, University of Konstanz, 78457 Konstanz, Germany}

%\date{\today}

\begin{abstract}

The transition dynamics of two-state systems with time-dependent energy levels, first considered by Landau, Zener, Majorana, and St\"uckelberg, is one of the basic models in quantum physics and has been used to describe various physical systems. We propose here a generalization of the Landau-Zener (LZ) problem characterized by distinct paths of the instantaneous eigenstates as the system evolves in time while keeping the instantaneous eigenenergies exactly as in the standard LZ model. We show that these paths play an essential role in the transition probability $P$ between the two states, and can lead to a substantial reduction of $P$, being possible even to achieve $P=0$ in an instructive extreme case, and also to large $P$ even in the absence of any anticrossing point. The partial LZ model can describe valley transition dynamics during charge and spin shuttling in semiconductor quantum dots.  

\end{abstract}

\maketitle

\section{Introduction}

The seminal works by Landau \cite{landau1932}, Zener \cite{zener1932}, St\"uckelberg \cite{Stueckelberg1932}, and Majorana \cite{majorana1932} revealed that a time-dependent drive in a quantum two-level system (TLS) leads to a transition between the states at an avoided crossing point, known as Landau-Zener (LZ) transition. Such transitions were investigated in a wide range of physical systems \cite{IVAKHNENKO20231}, such as graphene \cite{higuchi2017}, ultracold molecules \cite{mark2007}, quantum dots \cite{petta2010,ribeiro2013,cao2013}, classical resonators \cite{faust2012}, atomic qubits \cite{dupont2013}, Josephson junctions \cite{izmalkov2004}, and in quantum phase transitions due to topological defect formation in crystals \cite{pyka2013}. A TLS also defines a qubit, making it the basis for most quantum technologies, and rendering the LZ model highly relevant for the time-dependent control required in quantum computation and quantum information processing devices. 

\begin{figure}[th!]
\includegraphics[width=0.75\linewidth]{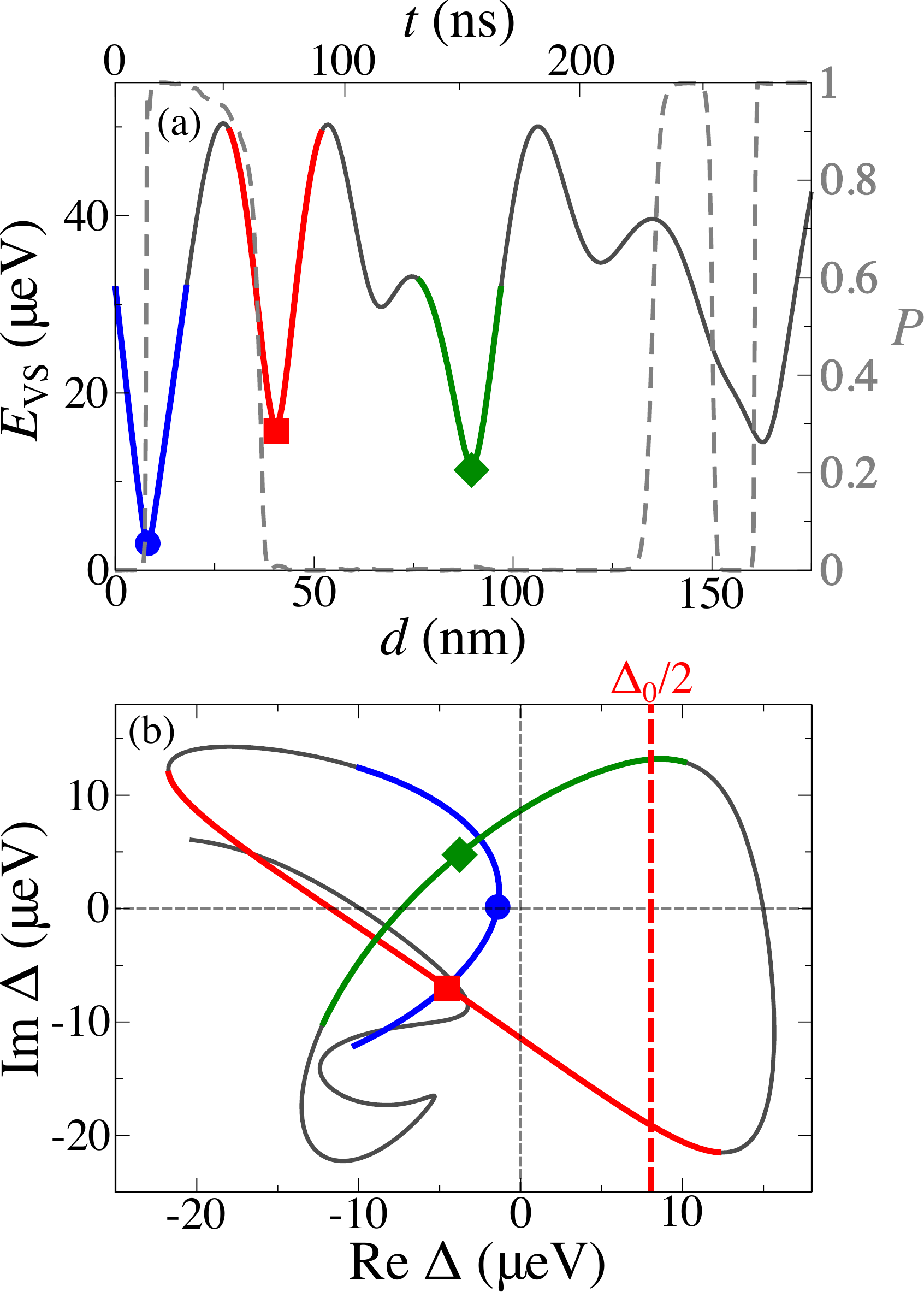}
\caption{Partial vs.\ Landau-Zener (LZ) transitions. (a) Energy splitting landscape as encountered, e.g., in electron shuttling with valley splitting $E_{\rm VS}=2|\Delta|^2$ (see Appendices~\ref{sec:appA} and \ref{sec:appB} for more details) \cite{Lima_2023,lima2024}. The splitting $E_{\rm VS}$ between two low-lying energy levels depends on the center position $d$ of a localized electron, where $d=vt$ during electron shuttling at velocity $v$.  Starting in the lower level at $t=0$,  $P$ denotes the excitation probability into the higher level.
(b) Concomitant trajectory of the coupling matrix element $\Delta$ in the complex plane. While the three highlighted minima of $E_{\rm VS}$ in (a) look similar, their different character can be witnessed in (b): LZ-like (red), Partial (blue), elliptic (green). For comparison, a standard LZ trajectory with energy gap $\Delta_0$ is shown (dashed red line).
\label{vs}}
\end{figure}

\begin{figure}[b]
\includegraphics[width=0.8\linewidth]{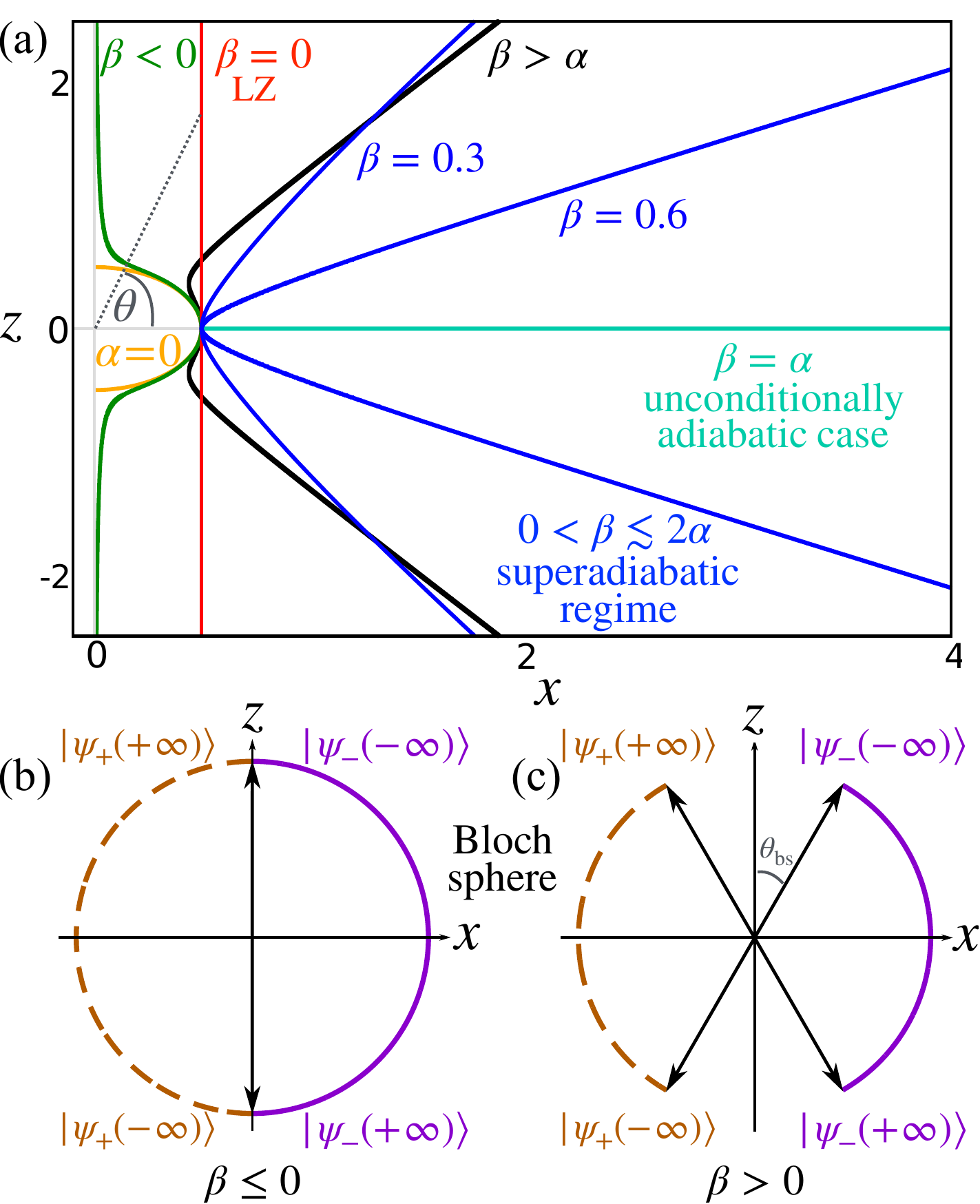}
\caption{(a) The paths in the parameter space as the system evolves in time for different values of $\alpha$ and $\beta$, where $z$ and $x$ are the diagonal and off-diagonal elements, respectively, of the Hamiltonian (\ref{eq:H}), which can be written as $H=-x\sigma_x - z\sigma_z$. We also define the angle $\theta$. (b) The projection of the instantaneous eigenvectors in the Bloch sphere and their path as the system evolves in time for the cases $\beta \leq 0$ and $\beta > 0$. $\theta_{\rm bs}=\pi/2-\theta$ is the angle between the eigenvectors and the $z$ axis in the Bloch sphere. \label{path}}
\end{figure}

The most general Hamiltonian for a two-level system can be written as $H=-\boldsymbol{r}\cdot \boldsymbol{\sigma}$, where $\boldsymbol{r}=(x,y,z)$ and $\boldsymbol{\sigma}$ are the Pauli matrices. In the LZ model, energy levels are swept past each other at constant level speed $\alpha$ and hybridize near the nominal crossing at $t=0$ to produce a fixed energy gap $\Delta_0$, with $x=\Delta_0/2$, $y=0$, and $z = \alpha t/2$. When the system evolves from $t=-\infty$ to $t=\infty$ the Hamiltonian curve is a straight line, as can be seen in the red (dashed) line in Fig.~\ref{vs}b and Fig.~\ref{path}a, and the probability of a transition between the two energy levels is given by the famous LZ formula $P_{\rm LZ} = \exp(-2\pi \Delta_0^2/\hbar\alpha)$.  Here, the energy gap $\Delta_0$ and level velocity $\alpha\ge 0$, and hence the transition probability $P_{\rm LZ}$, can be directly deduced from the instantaneous eigenvalues.  In this paper, we show that different Hamiltonian curves in the $xz$ plane can lead to different transition probabilities, even keeping the instantaneous eigenvalues unchanged.

A concrete physical example where this phenomenology is realized is the electron shuttling problem.
To address the need for long-distance coupling between remote qubits in quantum computing platforms, recent and ongoing experimental \cite{fujita2017,mills2019,jadot2021,noiri2022,Zwerver2023,vanriggelen2024,struck2024,desmet2024} and theoretical \cite{Ginzel2020,langrock2023,Bosco2024,losert2024} research has shown the possibility of transporting (shuttling) single electrons or holes along with their spin (qubit) degree of freedom across micrometers on a semiconductor chip. For the widely used Si/SiGe platform, electron shuttling along the disordered heterointerface is accompanied by a spatially varying energy splitting $E_{\rm VS}$ between the two quasi-degenerate lowest valley states in the Si conduction band (Fig.~\ref{vs}a) \cite{lima2024,volmer2024}. It is generally believed that a significant source of spin-qubit shuttling errors can be attributed to non-adiabatic LZ-type transitions due to a combination of different spin $g$ factors in the two valleys and the non-deterministic valley relaxation. Therefore, it is crucial to understand the probability of inter-valley transitions during the shuttling process. 

The Hamiltonian $H_v=-(\Delta^* \sigma_+ + \Delta \sigma_-)/2$ for this two-level valley system is given by $x=\text{Re}\; \Delta$, $y=\text{Im}\; \Delta$ and $z=0$, where $\Delta$ is the intervalley coupling matrix element. It turns out that the fundamental difference between $H_v$ and the LZ case is not the presence of the complex matrix element, since a simple rotation about the $x$ axis can transform $H_v$ into the LZ form with $y=0$.  However, $H_v$ is different because both $x$ and $y$ (and in the rotated case $x$ and $z$) are time-dependent. As a consequence, the instantaneous eigenstates trace out distinct Hamiltonian curves around the anticrossing point, in contrast to the straight line in the LZ case. This can be observed in Fig.~\ref{vs}b, where we show the paths around the anticrossing point in the $xy$ plane. Solving the time-dependent valley Hamiltonian we obtain the valley transition probability $P$ shown as the dashed gray line in Fig.~\ref{vs}b. Assuming that the system is initially in the ground state, $P = 0$ indicates that the system ends up in the ground state with certainty while  $P = 1$ stands for a guaranteed transition into the valley excited state. The valley transitions do not occur where one would expect based on the LZ model, which is a consequence of the different Hamiltonian curves. 

Generalizations of the LZ problem have been considered previously to describe various distinct systems. The initial and final times can be chosen finite rather than the asymptotic $t=\pm\infty$, leading to oscillating probabilities \cite{Vitanov1996}. With periodic driving, a sequence of transitions can take place where the St\"uckelberg phases of these transitions can interfere constructively or destructively, an effect known as a Landau-Zener-St\"uckelberg interference \cite{Shevchenko2010}. There are several examples of nonlinear LZ problems, where the sweep function $z(t)$ is no longer linear in $t$ \cite{Ashhab2022}. For the superlinear ($z(t)\sim t^\gamma$ with $\gamma>1$) and sublinear ($\gamma<1$) cases, the transition probability decreases and increases, respectively, compared to the LZ formula \cite{Vitanov1999}. Various sweep and gap functions were used, producing, e.g., an oscillating probability transition as a function of the parameters of the system \cite{rosen1932,hioe1984,suominem1992,Lehto_2016}. Another widely considered generalization of the LZ problem is the multilevel LZ problem \cite{pokrovsky2002,pokrovsky2004,shytov2004,garanin2008} where a large number of anticrossing points can be encountered as the system evolves in time, leading, e.g., to a distinct adiabaticity condition compared to the original LZ case.  We also mention here the many-body LZ problem, where the adiabatic regime is not reached even at very slow driving rates \cite{altland2008}. When we generalize the original LZ Hamiltonian, usually the problem is not exactly solvable. However, if certain conditions are satisfied, an approximate transition probability can be obtained, e.g., using the Dykhne-Davis-Pechukas (DDP) formula \cite{dykhne1960,dykhne1962,david1976}, which can be applied to a TLS with general sweep and gap functions. In a seminal work, Berry obtained that the transition probability between two quantum states is multiplied by an additional geometric factor \cite{berry1990}. However, this factor is equal to 1 for a Hamiltonian curve that lies in a plane through the origin. To the best of our knowledge, no previous work considered the influence of the geometry of the path $(x(t),z(t))$ on the transition probability. As we will show here, it plays a crucial role in the LZ transitions.

\section{Model}
We propose a generalized LZ model that produces the same instantaneous energy levels as the original LZ model but gives rise to transition probabilities that strongly deviate from the LZ formula.  The time-dependent Hamiltonian is constructed as $H(t)=E_+\ket{\psi_+}\bra{\psi_+}+E_-\ket{\psi_-}\bra{\psi_-}$, where $E_{\pm}=\pm \Omega_{\alpha}/2$, with $\Omega_{\alpha}=\sqrt{\Delta_0^2+\alpha^2t^2}$, are the LZ eigenvalues and $|\psi_{\pm}\rangle$ are new instantaneous eigenvectors which can deviate from the LZ case. We choose the real-valued
\begin{equation}
    |\psi_{\pm}\rangle = C_{\pm}(t)\left( \alpha t \mp \Omega_{\alpha}, \mp \beta t + \Omega_{\beta} \right)^T, 
\end{equation}
where $\Omega_{\beta} = \sqrt{\Delta_0^2+\beta^2t^2}$ and $C_{\pm}$ is a time-dependent normalization factor. The above definitions lead to,
\begin{equation}
    H=\frac{-\Omega_{\alpha}}{2(\Omega_{\alpha} \Omega_{\beta} \!-\! \alpha \beta t^2)}\left(\begin{array}{cc}
       (\alpha\Omega_{\beta}-\beta \Omega_{\alpha})t  & \Delta_0^2 \\
        \Delta_0^2 &   (\beta \Omega_{\alpha}-\alpha\Omega_{\beta})t
    \end{array}
    \right). \label{eq:H}
\end{equation}
This approach introduces the new parameter $\beta$. For $\beta=0$, we recover the LZ Hamiltonian.

Writing again $H(t)=-\boldsymbol{r}(t)\cdot \boldsymbol{\sigma}$, we visualize the trajectory $\boldsymbol{r}(t)=(x(t),0,z(t))$ for different values of $\alpha$ and $\beta$ in Fig.~\ref{path}(a). For $0<\beta<\alpha$ we have a hyperbola with curvature given by $\beta$, for the LZ case $\beta=0$ we find a straight line, and for $\beta=\alpha$ we have $z=0$, which corresponds to a straight back-and-forth trajectory on the positive $x$-axis. The case $\alpha=0$ relates to a finite trajectory where $\beta$ determines the maximum driving velocity. Here, despite the constant energy levels and consequently the absence of a crossing point, transitions will be possible, unlike in the LZ problem.
The parameter $\beta$ also changes how the projections of the instantaneous eigenvectors in the Bloch sphere rotate as the system evolves in time, as can be seen in Fig.~\ref{path}(b) and (c). From $t=-\infty$ to $t=+\infty$ the eigenvectors rotate and describe a closed circle in the Bloch sphere for $\beta \leq 0$, which includes the LZ case, while for $\beta > 0$ we find an open path.

\section{Results and discussion}
We numerically solve the time-dependent Schr\"odinger equation $i\hbar \partial_t \Psi(t) = H(t)\Psi(t)$ assuming that the system is initially in the ground state, $|\Psi(-t_0)\rangle = |\psi_-(-t_0)\rangle$ ($t_0 > 0$). We write the final state at $t=t_0$ as $|\Psi(t_0)\rangle = a|\psi_-(t_0)\rangle + b|\psi_+(t_0)\rangle$, where the probability of a transition is given by $P = |b|^2$.
\begin{figure}[h]
\includegraphics[width=\linewidth]{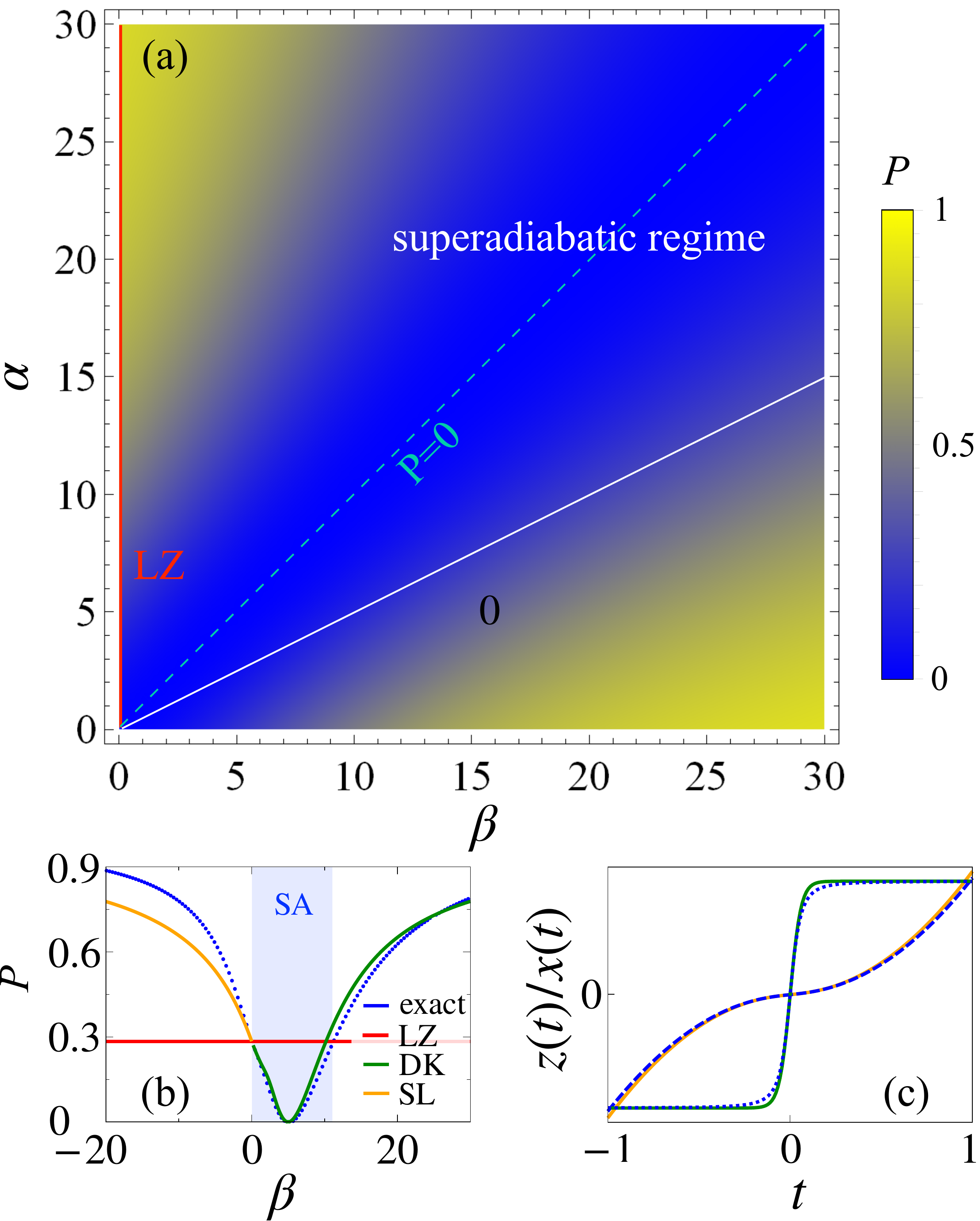}
\caption{Transition probability $P$ and superadiabatic (SA) behavior.
(a) $P$ as a function of $\alpha$ and $\beta$. The LZ case corresponds to $\beta=0$ (vertical red line). For increasing $\beta>0$, the decrease of $P$ indicates the SA behavior (SA regime bounded by red and white lines), until at $\beta=\alpha$, the unconditionally adiabatic case $P=0$ is reached.
(b) Our model yields SA behavior $P<P_{\rm LZ}$ in the range $0<\beta\lesssim 2\alpha$ (blue). Comparison of the transition probability $P(\beta)$ for $\alpha=5$ in a large range of $\beta$ obtained from the exact numerical solution of the time-dependent problem Eq.~(\ref{eq:H}) (blue dotted line), with various approximations. The LZ solution (red) only agrees with the exact solution for $\beta=0$ and $\beta\approx 2\alpha$. The DK formula Eq.~(\ref{eq:DK}) (green) constitutes a good approximation for $\beta>0$, while the SL model (orange) Eq.~(\ref{eq:SL}) provides a fair approximation for $\beta<0$.
(c) A comparison of DK (green) and SL (yellow) to the exact model with $\beta\le 0$ (dashed blue) and $\beta>0$ (dotted blue) in Hamiltonian parameter space explains the quality of approximations.
Note that $z(t)/x(t)$ has been rescaled for $\beta\le 0$.
\label{p}}
\end{figure}
In Fig.~\ref{p}(a) we plot the transition probability as a function of $\alpha$ and $\beta$ for a fixed energy gap $\Delta_0$. As expected, for $\beta = 0$, $P=P_{\rm LZ}$ is given by the LZ formula. As we increase the value of $\beta$, the transition probability decreases until we reach $P=0$ when $\beta = \alpha$. This suggests that the curvature of the path in the parameter space reduces the transition probability. However, if we keep increasing $\beta$, eventually we will have $P > P_{\rm LZ}$ when $\beta \gtrsim 2\alpha$, as can be seen in Fig.~\ref{p}(b), where the red line shows $P_{\rm LZ}$. For negative values of $\beta$ we also have $P > P_{\rm LZ}$, but in this case, we do not have the hyperbolic path in the $xz$ plane. It is important to remember that the eigenenergies do not depend on $\beta$. Thus,  these different behaviors all occur for the same energy landscape. 

The adiabaticity condition for the LZ problem follows from the adiabatic theorem  \cite{Born1928} and is given by $\Delta_0^2/\hbar \alpha \gg 1$. This condition ensures a very small transition probability $P_{\rm LZ}$ for large gaps and small level velocities. In our model, the adiabaticity condition is modified to $\Delta_0^2/\hbar |\alpha-\beta| \gg 1$ which explains $P<P_{\rm LZ}$ for $0<\beta<2\alpha$ and the unconditionally adiabatic dynamics for $\beta = \alpha$. We call $0<\beta\lesssim 2\alpha$ the \textit{superadiabatic} regime (see below for a detailed discussion of $\beta\approx 2\alpha$). This regime is desired in various physical systems. For instance, as already mentioned, during spin shuttling in SiGe quantum dots, superadiabatic dynamics can occur,  suppressing inter-valley transitions, and, in turn, increasing the spin coherence, and consequently enhancing the performance and scalability of silicon spin qubits. Procedures to achieve adiabaticity through a fast route, known as shortcuts to adiabaticity (STA), were widely proposed previously \cite{STA}. Our approach is different. We are describing systems that have an adiabatic time evolution where none would be expected from a standard Landau-Zener model, while STA achieves a speedup by giving up adiabatic evolution (in a controlled way).

Fig.~\ref{p}a also shows that the transition probability is symmetric when we interchange $\alpha$ and $\beta$, even though the Hamiltonian does not have this symmetry. This also means that when $\alpha = 0$, the transition probability is given by the LZ formula, with $\beta$ replacing $\alpha$. In this case, we still have a time-dependent Hamiltonian, but the eigenenergies are constant in time. All these results reveal that the path in the $xz$ plane (Fig.~\ref{path}(a)) plays a crucial role and should be considered to correctly predict the transition probability.

Further insight is gained from the change of the angle, 
\begin{equation}
\theta(t) = \arctan \left(\frac{(\alpha \Omega_{\beta}(t) - \beta \Omega_{\alpha}(t))t}{\Delta_0^2}\right),
\end{equation}
in the $xz$ plane  over time (Fig.~\ref{path}).
Numerically, we find that the transition probability depends on the instantaneous angular velocity at the crossing point 
$\dot{\theta}(0) = |\alpha-\beta|/\Delta_0$, and on the range of $\theta(t)$ when the system evolves from $t=-\infty$ to $t=\infty$. The extreme values of $\theta$ are $\pm \arctan((\alpha^2-\beta^2)/2\alpha \beta)$ for $\beta > 0$ and  $\pm \pi/2$ for $\beta \leq 0$. Therefore, the increase of $P$ for negative values of $\beta$ is due to the increase of $\dot{\theta}(0)$. Moreover, the unconditionally adiabatic case at $\beta = \alpha$ is a consequence of a static $\theta\equiv 0$, which implies $\dot{\theta}(0) =0$. We note that $\dot{\theta}(0)$ is symmetric around $\alpha=\beta$, but $P$ is not, as seen in Fig.~\ref{p} (b). This is because $\theta$ changes by a wider range for $\beta < \alpha$ compared to $\beta > \alpha$. Thus, we find a higher transition probability for $\beta < 0$, which confirms that the curvature of the Hamiltonian curve reduces the transition probability.
This also explains that the upper bound of the superadiabatic regime is not exactly at $\beta=2\alpha$.

We also find that $P$ is symmetric under interchange of $\alpha$ and $\beta$ because $\theta$ also has this symmetry. In fact, the interchange of $\alpha$ and $\beta$ changes the sign of $\theta$, but it does not influence $P$. As a consequence, we can use simpler approaches to map the transition probabilities of more complex systems. For example, in the case $\alpha = 0$ and $\beta > 0$, even though we have an analytical solution, it is given as a linear combination of distinct confluent Heun functions and their first derivatives. The asymptotic behavior of this solution cannot be easily accessed to obtain an analytical formula for the transition probability. However, since this case has the same $\theta$ as in the case $\beta=0$ and $\alpha > 0$, which is the original LZ problem, the transition probability is given by the LZ formula.

It is important to mention that two distinct two-level systems with the same $\theta$ do not necessarily have the same $P$. The LZ problem with drive velocity $\alpha_1$ and energy gap $\Delta_1$ has, e.g., the same $\theta$ as the LZ problem with $\alpha_2 = 2\alpha_1$ and $\Delta_2 = 2\Delta_1$, but these two systems have different $P_{\rm LZ}$. This is because the LZ formula depends on $\Delta_0^2/\alpha$. To have the same $P_{\rm LZ}$, when increasing the energy gap by a factor of two, we have to increase $\alpha$ by a factor of four. Therefore, we can say that two distinct two-level systems have the same transition probabilities if they have the same $\theta$ and $\Delta_0$.

Another way to understand our results is by considering the angle $\theta_{\rm bs}$ between the instantaneous eigenvectors and the $z$-axis when we project them in the Bloch sphere, shown in Fig.~\ref{path}c. It turns out that $\theta_{\rm bs} = \pi/2 - \theta$ and thus all previous discussions about the influence of $\theta$ on the transition probability  also apply to $\theta_{\rm bs}$. We add that the LZ transition is a consequence of the rotation of the instantaneous eigenvectors in the Bloch sphere.  These transitions take place even without the presence of an avoided crossing point, which we have shown in the case $\alpha = 0$.

While we are not aware of an analytical solution to our model, one can attempt to use the DDP formula to obtain an approximate transition probability, since the necessary conditions for this are satisfied. The DDP formula is written in terms of the zeros of the function $\mathcal{E}(t)=\sqrt{z^2(t)+x^2(t)}$, where $t$ is treated as a complex variable. Even though the functions $z(t)$ and $x(t)$ in our model are different from the  LZ problem, the function $\mathcal{E}(t)$ is the same in both cases. As a consequence, the DDP approach fails in our model, since it predicts a transition probability equal to the LZ case. A possible reason is that the zero $t_c=i\Delta_0/\alpha$ of $\mathcal{E}(t)$ also yields $H(t_c)=0$, which suggests the existence of additional conditions for the applicability of the DDP formula. Another failure of the DDP approach in a nonlinear LZ model was reported recently \cite{Ashhab2022}. These results call into question the validity of the DDP formula, which is a topic that deserves to be addressed in future works. 

Another possibility to approximate $P$ based on our previous discussion consists in using a simpler model with the same (or similar) $\theta$ to map our system. We plot $\tan \theta = z(t)/x(t)$ in Fig.~\ref{p}c for $\beta >0$ and $\beta \leq 0$. For $\beta >0$, we can find a model with a similar $\theta$ by fitting the curve with the function $a\tanh(bt)$, where $a=(\alpha^2-\beta^2)\Delta_0/(2\alpha\beta)$ is obtained by taking the limit $t \rightarrow \pm \infty$ in $\tan \theta$ and $b=2\alpha \beta /(\alpha+\beta)$ ensures that the instantaneous angular velocity at the anticrossing point is the same in both cases. Therefore, we can approximate our system for $\beta >0$ using $z(t)=a\tanh(bt)$ and $x=\Delta_0$. This model has an analytical solution and was first solved by  Demkov and Kunike (DK) \cite{dk,suominem1992,requist2005}. The transition probability is given by,
\begin{equation}
    P_{\rm DK} \approx \frac{\sinh^2(\pi a/b)}{\sinh^2(\pi\sqrt{a^2+\Delta_0^2}/b)} .
    \label{eq:DK}
\end{equation}
We plot $P_{\rm DK}$ in Fig.~\ref{p}(b) (green curve), which shows a good agreement with the numerical results. We have that $P_{\rm DK}\rightarrow 1$ when $\beta \rightarrow 0$, which means that this approximation breaks down for very small values of $\beta$. We found a reasonable agreement between $P_{\rm DK}$ and our numerical results when $\beta/\alpha \gtrsim  10^{-4}$.

The case $\beta \leq 0$ can be approximated by a model with $z(t)=(\alpha-\beta)t\sqrt{1-\beta t^2}$ and $x=\Delta_0$ (Fig.~\ref{p}c). This is the superlinear (SL) LZ model \cite{Vitanov1999} which has no exact solution. However, the DDP approach yields,
\begin{equation}
    P_{\rm SL}\approx e^{-\pi\frac{\Delta_0^2}{2(\alpha-\beta)}}.
    \label{eq:SL}
\end{equation}
In this case, we find a good agreement with the numerically exact results for $|\beta|$ not too large, which can be seen in the orange curve in Fig.~\ref{p}b. The deviation for large $|\beta|$ is due to the combination of two approximations. We emphasize that even though we are using the DK and SL models as approximations, these models are completely inequivalent to ours.

\section{Consequences for shuttling} 
Shuttling an electron over a distance of $d=200\, {\rm nm}$ with a constant velocity of 1~mm/s in the system shown in Fig.~\ref{vs} leads to the valley excitation probability $P=0.67$, i.e., shuttling fidelity of $1-P=33\%$. Based on the standard LZ model, one can enhance the fidelity by moving faster in regions with large $E_{\rm VS}$ and slowly when $E_{\rm VS}$ is small. This allows, e.g., for a higher fidelity of 57$\%$ while moving at a larger average velocity of 0.1~m/s. However, based on our results, an even better strategy is to limit the angular velocity $\dot{\theta}$. We obtained numerically that the system is in the adiabatic regime when $\dot{\theta}$ has an order of magnitude around $10^8$~rad/s or lower. Therefore, tuning the shuttling velocity in such way that $\dot{\theta} = 10^8$~rad/s at all times, which means that we are moving fast (slowly) when $\theta$ changes slowly (fast), a very large shuttling fidelity, exceeding 99.99$\%$, can be reached while moving at an even larger average velocity of 0.5~m/s, which constitutes a drastic improvement compared with the previously mentioned strategies.

\section{Conclusions}
We have proposed a generalization of the LZ model that gives rise to a superadiabatic (SA) regime while maintaining the same instantaneous eigenenergies as in the LZ case. This model is relevant, as it can be used to describe, e.g., the valley transitions during electron spin shuttling in SiGe quantum dots. The presented approach reveals how the Hamiltonian curve in the $xz$ plane determines the transition probability $P$. We have described the SA regime, where $P$ is significantly lower than in the LZ case, and the unconditionally adiabatic case, where $P=0$ no matter the driving velocity and energy gap. This offers a new path towards the engineering of driven dynamics with reduced excitation probability in the SA regime. We have also found that LZ transitions are possible even without the presence of any avoided crossing point in the energy levels, revealing that transitions are a consequence of the rotation of the eigenvectors in the Bloch sphere. Future extensions may include dissipative systems which have been widely investigated in the context of LZ transitions \cite{Wubs2006,Saito2007,Nalbach2009,Huang2018,Zheng2021}. The effects of unavoidable dissipation in the extended model and SA regime deserve further investigation. Quantum control schemes for quantum systems coupled to a thermal bath can also be applied to our model \cite{PhysRevA.101.052102,PhysRevResearch.3.013064}, particularly to the SA regime. Future work may also explore the influence of the topology of the closed paths of the instantaneous eigenvectors in the adiabatic regime.

\begin{acknowledgments}
This work has been funded by the Federal Ministry of Education and Research (Germany), Grant No.~13N15657 (QUASAR),
and by Army Research Office (US) grant W911NF-15-1-0149.
\end{acknowledgments}

\appendix

\section{Valley splitting in a silicon quantum dot}
\label{sec:appA}

%\begin{figure}[h]
%\includegraphics[width=0.7\linewidth]{system}
%caption{\small Schematic diagram of the SiGe/Si/SiGe heterostructure containing the quantum dot. The device has six regions: the Si well, the top and bottom interfaces, the lower and upper SiGe barriers and a insulator layer. We take into account the interface roughness, which means that the interface width changes randomly as function of $x$ and $y$, as be seen in the red and blue surfaces. We consider that the potential is infinite in the insulator region.} \label{system}
%\end{figure}

Within the effective mass theory,  the two low-lying valley states of a silicon quantum dot in a SiGe/Si/SiGe heterostructure grown in the $\hat{z}$ direction can be written as \cite{PhysRevB.75.115318}
\begin{equation}
|\pm z\rangle = \Psi_{xyz}(\textbf{r})e^{\pm ik_0z}u_{\pm z}(\textbf{r}),
\end{equation}
where $ \Psi_{xyz}(\textbf{r})$ is the envelope function, $u_{\pm z}(\textbf{r})$ are the periodic parts of the Bloch wavefunctions at the two conduction band minima (valleys) of silicon, $k_0 = 0.82(2\pi /a_0)$ is the Bloch wavenumber at the two valleys, and $a_0 = 0.543$~nm is the length of the Si cubic unit cell. 

We model the confinement potential due to the SiGe barriers by a sum of delta functions at the location of each Ge atom \cite{Lima_2023,lima2024},
\begin{equation}
U(x,y,z)=\lambda \sum_i \delta(x-x_i) \delta(y-y_i) \delta(z-z_i),
\label{U}
\end{equation}
where $\lambda = 10$~meV$\cdot$nm$^3$ is a fixed parameter of the model and $i$ labels the Ge atoms. We assume that the Ge atoms are distributed uniformly in the $\hat{x}$ and $\hat{y}$ directions and we model the smooth interface in $\hat{z}$ direction by a probability distribution function given by a hyperbolic tangent function.

The intervalley coupling is given by
\begin{eqnarray}
\Delta(\textbf{r}) &=& \langle +z |U(x,y,z)|-z\rangle \nonumber \\
&=&C_0\int e^{-2ik_0z}U(x,y,z)|\Psi_{x,y,z}(\textbf{r})|^2 d^3x,
\end{eqnarray}
where $C_0 = -0.2607$ results from the periodic parts of the Bloch wavefunctions \cite{PhysRevB.84.155320,PhysRevResearch.2.043180}. 
The total valley splitting is then given by
\begin{eqnarray}
E_{\rm VS}(\textbf{r}) = 2 |\Delta(\textbf{r})|.
\end{eqnarray}

Due to the random distribution of the Ge atoms at the SiGe/Si interfaces, the so-called alloy disorder, the intervalley coupling, and consequently the valley splitting, depends on the location of the quantum dot, which gives rise to the random oscillations of the valley splitting in Fig.~1 (a) and the random path in the complex plane in Fig.~1 (b) during charge/spin shuttling. 

The valley splitting landscape $E_{\rm VS}(d)$ in Fig.~1 was obtained assuming a SiGe/Si/SiGe heterostructure with a Si well of 10~nm thickness and SiGe barriers with randomly distributed 30$\%$ Ge. Also, a Si/SiGe interface width of 22 monolayers, a quantum dot radius of 21~nm, and an electric field in the $z$ direction of 20 MV/m were used \cite{lima2024,Lima_2023}.

\section{Valley shuttling fidelity}
\label{sec:appB}

The Hamiltonian for the two valley states can be written as
\begin{eqnarray}
H = \left( \begin{array}{cc}
 0 & \Delta(\textbf{r})  \\ 
\Delta(\textbf{r})^* & 0 
\end{array} \right).
\label{eq:valleyH}
\end{eqnarray}
Using the polar decomposition $\Delta=|\Delta|e^{i\theta}$, the two valley eigenstates can be written as 
\begin{eqnarray}
\alpha_{\pm}(\textbf{r}) = \frac{1}{\sqrt{2}}\left( 
\begin{array}{c}
e^{i\theta(\textbf{r})/2} \\
\pm e^{-i\theta(\textbf{r})/2}
\end{array}\right).
\end{eqnarray}
These are the instantaneous eigenstates of Eq.~\eqref{eq:valleyH} with corresponding instantaneous eigenenergies,
\begin{equation}
E_{\pm}(\textbf{r}) = \pm |\Delta(\textbf{r})|,
\end{equation}
 giving rise to the valley splitting $E_{\rm VS}(\textbf{r}) = E_{+}(\textbf{r}) -E_{-}(\textbf{r})$ in Eq.~(4).
For a  quantum dot moving in the $\hat{\textbf{x}}$ direction at the (shuttling) velocity $v$, the Schr\"odinger equation,
\begin{equation}
i\hbar \partial_t \psi(t) = H(t)\psi(t),
\end{equation}
can be rewritten employing a change of variables, $\textbf{r} = v \hat{\textbf{x}} t$, and thus $x=vt$, as
\begin{equation}
i\hbar v \partial_x \psi(x) = H(x)\psi(x).
\label{hx}
\end{equation} 
We solve Eq.~(\ref{hx}) numerically considering that initially the electron is in the valley ground state $\alpha_-$ and write
\begin{equation}
\ket{\psi (x)} = a(x)\ket{\alpha_-(x)}+b(x)\ket{\alpha_+(x)}.
\end{equation}
The probability of a valley transition is then given by
\begin{equation}
P(x) = |\bra{\alpha_+}(x)\ket{\psi(x)}|^2 = |b(x)|^2.
\end{equation}

% Create the reference section using BibTeX:
\bibliography{ref.bib}

\end{document}